\documentclass[fleqn,12pt,twoside]{article}
\usepackage{espcrc1}

\usepackage{graphicx}
\usepackage[figuresright]{rotating}

\newcommand{\AmS}{{\protect\the\textfont2
  A\kern-.1667em\lower.5ex\hbox{M}\kern-.125emS}}

\title{Correlations, Fluctuations, and Flow Measurements from
the STAR Experiment} 

\author{R. L. Ray\address{Department of Physics, The University of Texas 
        at Austin, \\ 
        Austin, Texas 78712 U.S.A.} for the STAR Collaboration\footnote{
For the full author list and acknowledgements, see Appendix ``Collaborations''
of this volume.}}
       
\begin{document}

\maketitle

\begin{abstract}
New measurements of short-range and long-range two-particle correlations,
azimuthal an\-i\-so\-tro\-py, and event-by-event fluctuations from the STAR
experiment for $\sqrt{s_{NN}}= 130$ and 200~GeV Au+Au collisions
are summarized.
Striking evidence is presented for large, non-statistical fluctuations in
mean transverse momentum.
Descriptions of the data in terms of phenomenological source
function models are also presented. 
\end{abstract}

\section{Introduction}

Determining whether or not color deconfinement and chiral symmetry restoration
occur during relativistic heavy ion collisions at RHIC requires a multi-prong
effort involving experiment, phenomenology, and theory.
Experimentally, measurements of inclusive
spectra and correlation distributions for hadrons and leptons
are needed to study quantum interference, final state interactions,
and non-statistical fluctuations which produce correlations
evident in the observed momentum distributions.
Detailed studies of non-statistical event-by-event 
fluctuations~\cite{ebye} by way of correlation measurements
are possible at RHIC, especially with the large
acceptance STAR detector~\cite{star}.
To help bridge the gap between
experiment and theory, phenomenological source function models are being used
by both experimentalists and theorists to describe the data.

This paper summarizes recent correlation measurements and phenomenological
source model descriptions from the STAR experiment for soft
hadron production in Au+Au collisions at $\sqrt{s_{NN}} =130$ and 200~GeV
at RHIC.  The data include results on Hanbury Brown and Twiss (HBT)
pion interferometry, pion phase space densities,
pion-kaon correlations, long-range correlations in pseudorapidity space,
azimuthal anisotropy, and event-by-event fluctuations in mean transverse
momentum.  Data presented here were obtained with the STAR
Time Projection Chamber (TPC) using the Level~0 trigger detectors
CTB (Central Trigger Barrel) and
ZDC (Zero-Degree Calorimeter)~\cite{star} and should
be regarded as preliminary. 

\section{Source Function Phenomenology}

Models of the space-time and momentum distribution of the hadronic freeze-out
region in relativistic heavy ion collisions
typically have the following form (for bosons)~\cite{ssh}:
\begin{equation}
S(x,p) = \frac{m_T \cosh (y-\eta)}{(2\pi\hbar)^3} \left\{ \exp \left(
\frac{p \cdot u(x) - \mu}{T} \right) - 1 \right\}^{-1} F(r,\varphi,\eta,\tau),
\label{Eq1source}
\end{equation}
where $x$ [with components $r,\varphi,\eta$ (source rapidity)
and $\tau$] is the space-time coordinate,
$p$ (with components $m_T,y,\phi$) is the on-shell momentum coordinate,
$m_T$ is the transverse mass, $u(x)$ represents the collective velocity field
which accounts for longitudinal (Bjorken) and transverse expansions,
$\mu$ is the chemical potential, $T$ is the local temperature, and $F$ is
an empirical modulation function.  The latter
may depend on the radial, azimuthal,
rapidity and longitudinal proper time ($\tau = \sqrt{t^2 - z^2}$) coordinates
of the source. 
A special case of the general emission function in Eq.~(\ref{Eq1source}) is
referred to as the ``blast-wave''
model.  This model assumes longitudinal boost invariance, finite
freeze-out duration with respect to
a fixed longitudinal proper time $\tau_0$, and integration over the observed
rapidity $y$.  The blast-wave model has been used to describe some of the
STAR data taken in the first two RHIC runs as discussed below.

\section{HBT Pion Interferometry}


Preliminary results for three-dimensional
Pratt-Bertsch~\cite{prattb} ``out-side-long''
HBT analyses of the recent $\sqrt{s_{NN}} = 200$~GeV Au+Au
minimum bias collision data from STAR for charged pions (see
contribution to these proceedings by M. Lopez-Noriega) are very similar to
the STAR~\cite{starhbt} and PHENIX~\cite{phenixhbt}
results at 130~GeV. 
The
measured three-dimensional correlations were fitted with the standard
gaussian form,
\begin{equation}
C(q_o , q_s , q_{\ell} ) = 1 + \lambda \exp [ -R_o^2 q_o^2 - R_s^2 q_s^2
- R_{\ell}^2 q_{\ell}^2 ]
\end{equation}
where the relative momentum vector was evaluated in the 
longitudinal co-moving source (LCMS) frame and decomposed into components
$q_{\ell}$ (parallel to beam direction), $q_o$ (parallel to pair transverse
momentum) and $q_s$ (orthogonal to each).  Standard cuts and corrections
for track splitting, two-track merging in the STAR TPC, and
Coulomb corrections were included as in Ref.~\cite{starhbt}. 
The correlation lengths\footnote{
The conjugate lengths of the relative momentum
correlations characterize the homogeneity
scale of the emitting source.  It is customary to refer to these
lengths as ``radii.''  For expanding sources
additional analysis is required to determine the
total source geometry.} or ``radii,'' $R_o,R_s,R_{\ell}$,
at $\sqrt{s_{NN}} = 200$~GeV
are similar in magnitude and have the same
centrality and transverse mass
($m_T$) dependences as the results at $\sqrt{s_{NN}} = 130$~GeV. 
$R_{\ell}$ adheres to the
Makhlin-Sinyukov~\cite{sinyukov}
dependence on $m_T$ expected for Bjorken expansion
with corresponding source lifetimes which range from approximately
8~fm/c for peripheral collisions to 10~fm/c for central events.  The
ratio $R_o/R_s$ for the 200~GeV data is approximately unity in agreement with
results reported at 130 GeV~\cite{starhbt}.  This result is not understood
theoretically but can be described with a
brief emission duration ($\Delta\tau$)
and sharp radial cut-off of the source geometry.

%

\begin{figure}[htb]
\begin{minipage}[t]{65mm}
\includegraphics*[scale=0.4]{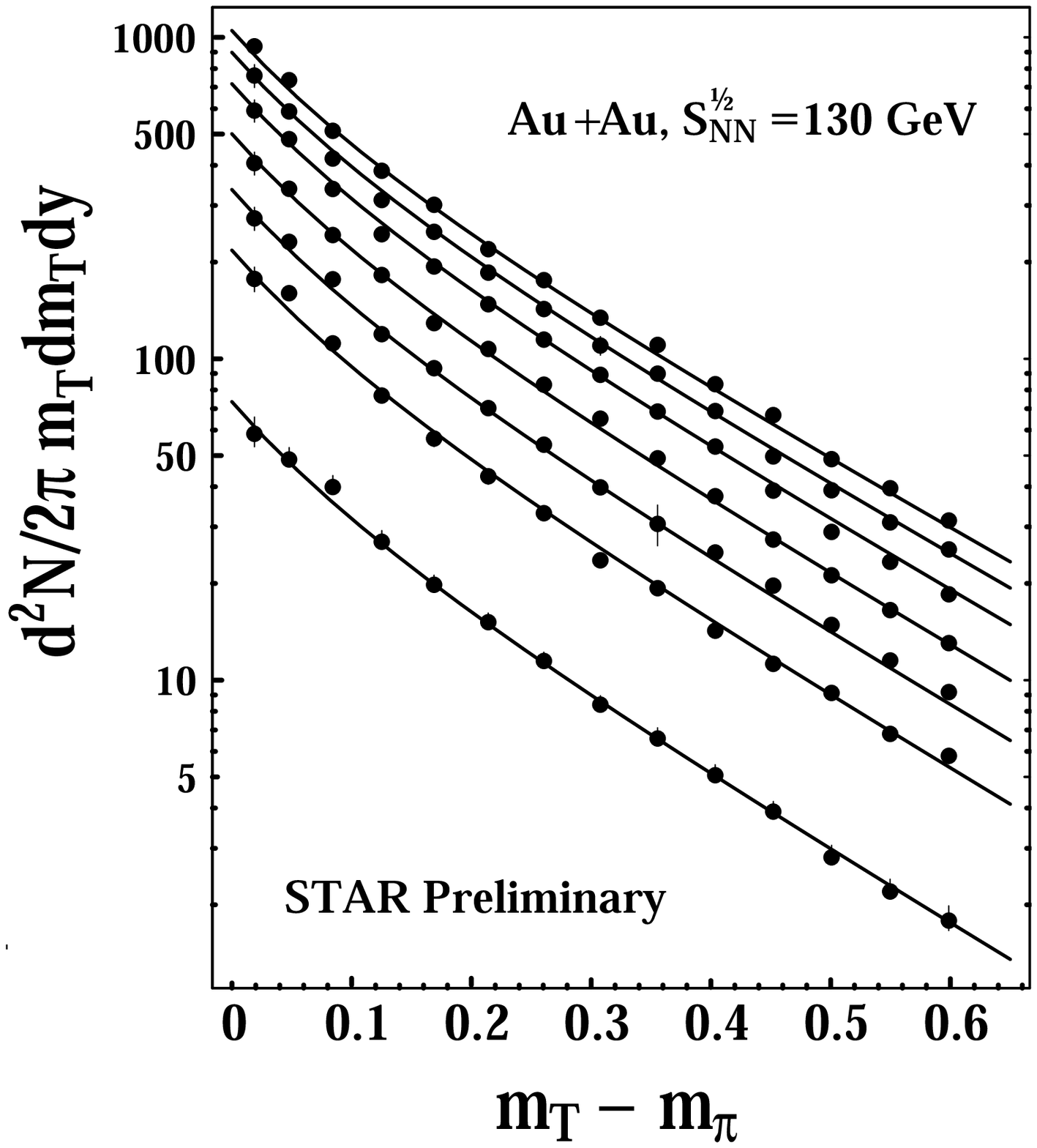}
\vspace{-10mm}
\caption{Inclusive $\pi^-$ transverse mass spectra in units of GeV
for seven centralities with BE distribution model
fits (solid curves).}
\label{Fig:Pionspectra}
\end{minipage}
\hspace{\fill}
\begin{minipage}[t]{90mm}
\includegraphics*[scale=0.55]{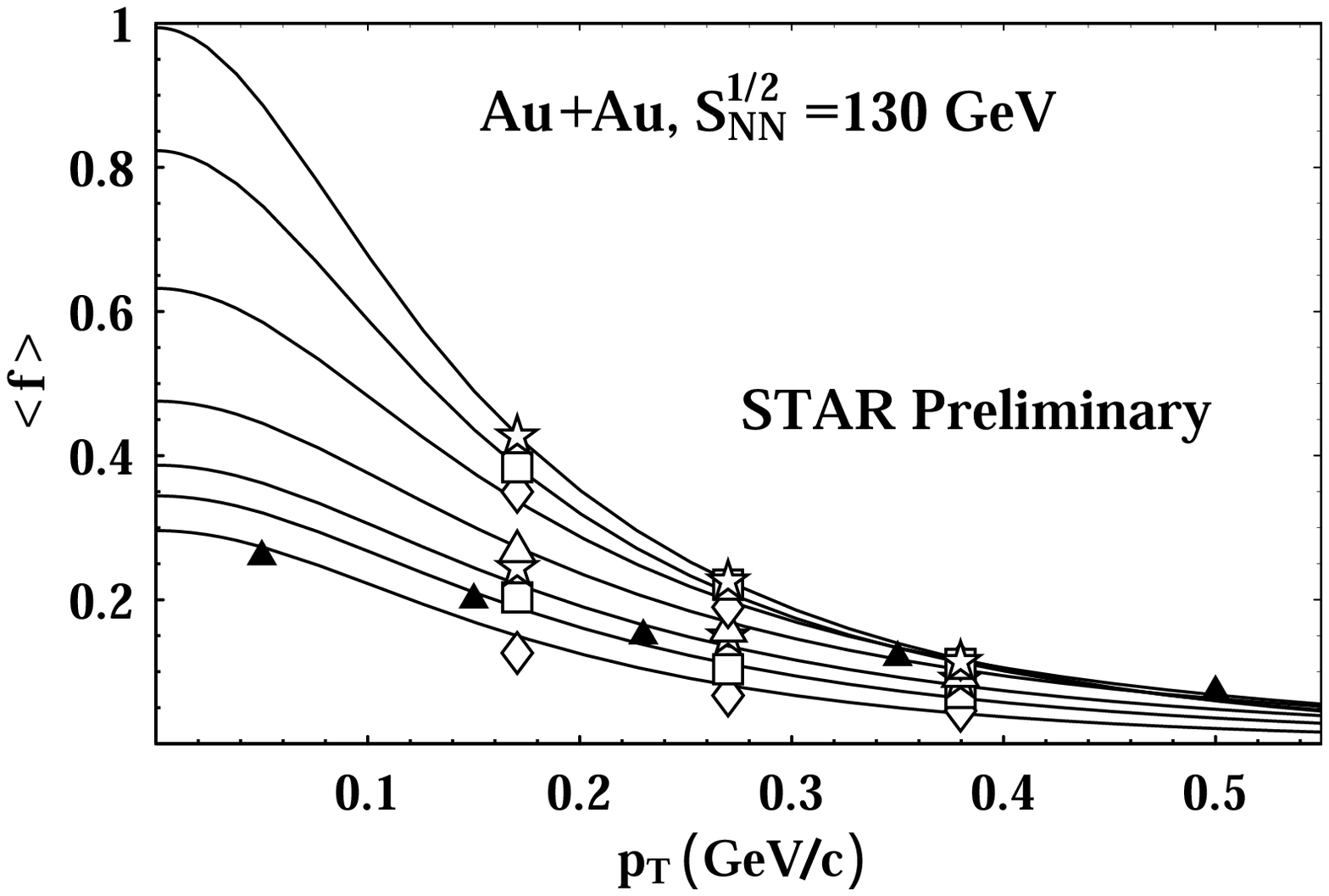}
\vspace{-10mm}
\caption{Average $\pi^-$ PSD versus $p_T$
for seven centralities
from STAR and from NA49 for top central events (solid triangles)
with BE distribution model fits
(solid curves) (see text).}
\label{Fig:PionPSD}
\end{minipage}
\end{figure}

Another important aspect of the pion emission source which can be inferred
from the HBT results is the phase space density (PSD).
Bertsch~\cite{bertschpsd} and Ferenc {\it et al.}~\cite{ferenc}
have shown that the pion
transverse momentum ($p_T$) distribution together with the volume of the
two-pion HBT correlation function can be used to determine the spatial average
of the pion PSD, $\langle {\rm f} \rangle$. 
This method was applied to the
minimum bias $\sqrt{s_{NN}} = 130$~GeV Au+Au collision data from STAR
for negative pions for seven centrality bins
[defined by the percent ranges 0-5 (most central), 5-10, 10-20, 20-30, 30-40,
40-50, and 50-80 (peripheral) of the total reaction cross section].
The $m_T$ distribution data,
shown in Fig.~\ref{Fig:Pionspectra},
were fitted with a Bose-Einstein (BE) distribution,
$\{A/[ \exp (m_T/T_{eff}) - 1]\}$, with effective temperatures
approximately 200~MeV (solid curves). 
The spectra in Fig.~\ref{Fig:Pionspectra} increase in magnitude
monotonically with centrality.
The volumes of the three-dimensional Pratt-Bertsch HBT correlation
functions for these seven centralities were obtained
by direct summation of the binned data.
The resulting estimates of $\langle {\rm f} \rangle$ for three
ranges of $p_T$ (defined by the limits 0.125, 0.225, 0.325, and 0.45~GeV/c)
are shown in
Fig.~\ref{Fig:PionPSD}.
For comparison, similar estimates for central Pb+Pb
collisions at the SPS from experiment NA49~\cite{ferenc}
are indicated by the solid triangles
in Fig.~\ref{Fig:PionPSD}.  The results show that $\langle {\rm f} \rangle$
increases with $\sqrt{s_{NN}}$
from SPS to RHIC and monotonically increases with centrality at RHIC energies.
Apparently the universal pion PSD
suggested by Ferenc~\cite{ferenc} for AGS and SPS energies is not valid at
RHIC energies. 

An interesting prediction follows from the
source function model of Tom\'{a}\v{s}ik and Heinz~\cite{ssh},
which includes explicit longitudinal expansion effects,
when fitted to the negative pion $m_T$ spectra in Fig.~\ref{Fig:Pionspectra}
and the PSD values in
Fig.~\ref{Fig:PionPSD}
(solid curves).  The model, when extrapolated to low $p_T$,
predicts large values of $\langle {\rm f} \rangle \approx 1$ for central
events, suggesting that significant multiparticle Bose-Einstein effects may be
present at low $p_T$ at RHIC.

\section{Non-Identical Two-Particle Short-Range Correlations}

Lednick\'{y} {\it et al.}~\cite{lednicky}
pointed out that measurements of short-range relative momentum
correlations in the co-moving pair frame of non-identical
particles could
be used to constrain dynamical models of the emission source, in
particular the average space-time differences between emission
regions for different particle species.\footnote{This analysis method cannot
simultaneously determine spatial and temporal differences.} 
%
For example,
for $\pi^+$K$^-$, the attractive Coulomb interaction produces correlations
greater than unity.  Correlations obtained for pairs where the pion speed
is selected to be greater than the kaon's are denoted
$C^{\pi K}_+$ and vice versa
for $C^{\pi K}_-$.  If the pion and kaon emission
times are the same, for example, observation
of $C^{\pi K}_+/C^{\pi K}_- > 1$ implies that
the average $\pi^+$ emission radius is smaller
than that of the $K^-$ and vice versa if $C^{\pi K}_+/C^{\pi K}_- < 1$.
Similar arguments apply to the other charge sign combinations.

Three-dimensional Pratt-Bertsch~\cite{prattb}
$\pi^{\pm} K^{\pm}$ correlation ratios for the top 12\% central events
from the 130~GeV Au+Au collision data from STAR were measured and analyzed
(see contribution to these proceedings by F. Retiere).  Transverse velocity
magnitudes ($\beta$) were restricted to the range from $\sim 0.7-0.75c$
due to limitations imposed by dE/dx based particle identification with the TPC.
Due to the inherent ambiguity in the analysis
the data are consistent with pion emission which in one limit
occurs at relatively smaller
transverse position ($\sim 4.1$~fm in the lab frame) than the kaons, assuming
equal emission times, or in another limit occurs
an average of 5.4~fm/$c$ later
than kaons assuming emission at equal transverse positions. 

These $\pi^{\pm} K^{\pm}$ correlation ratios can be understood in terms of the
``blast-wave'' model discussed above.
The selection of $\beta \approx 0.72c$ particles together with
the brief emission duration and sharp radial cut-off of the emission
source geometry in the blast-wave
model results in the emission zone being a thin shell near the outer edge
of the source.  Thermal smearing effects, which are larger for lighter
mass particles (pions), conspire with the latter effect
to cause the average pion emission radius to be smaller than
the kaon's.
The parameters of the blast-wave model which describe these data [$T$ = 110~MeV,
$\langle \rho \rangle = 0.6$ (average transverse flow rapidity), R = 13~fm,
$\Delta \tau = 1.5$~fm/$c$ and $\tau_0 \approx 10$~fm/$c$],
are consistent with those which describe the $p_T$ distribution shapes,
pion HBT radii, and $R_o/R_s$ ratio.

\section{Long-Range Correlations in Pseudorapidity Space}

Two-particle momentum space correlations in general
are six-dimensional objects.  Depending on the nature of the dynamics to be
studied and possible limitations of statistics,
projections onto fewer dimensions, usually
one or two, are typical in most analyses. 
For the STAR data presented,
{\it charge independent} (CI) and
{\it charge dependent} (CD) combinations of charged-sign pairs were formed. 
The CI correlations were obtained by
summing over all pairs regardless of charge sign. 
The CD correlations are defined by
$C({\rm CD}) = ( C_{++} + C_{--} ) -
( C_{+-} + C_{-+} )$, where $C_{\pm\pm}$ are charge-pair specific
correlations.
 
\begin{figure}[htb]
\begin{minipage}[t]{80mm}
\hspace*{5mm}
\includegraphics*[scale=0.35]{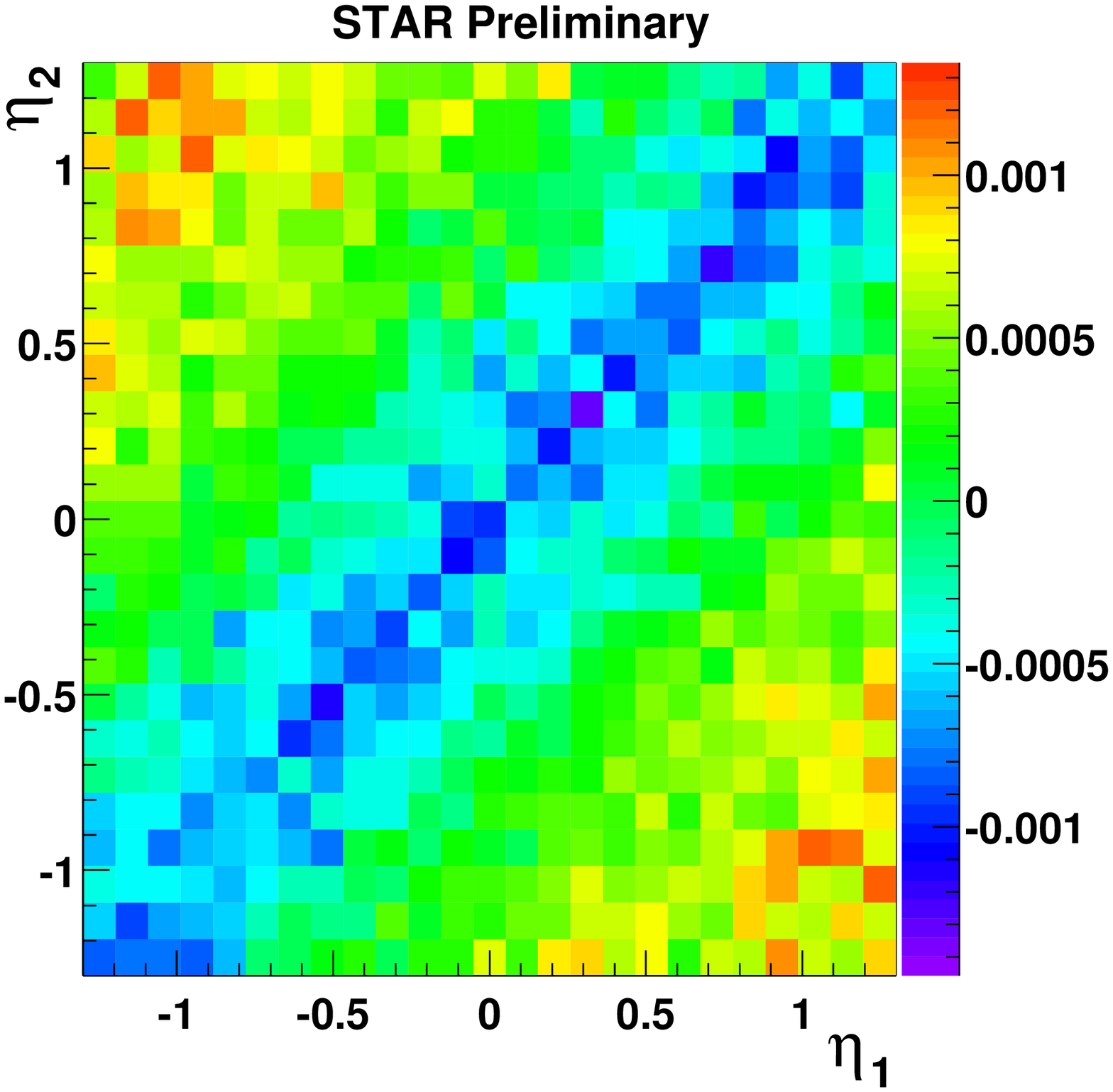}
\vspace{-10mm}
\caption{Two-dimensional $\eta_1 \otimes \eta_2$ charge dependent correlations
for $\sqrt{s_{NN}} = 130$~GeV Au+Au collisions.}
\label{Fig:etaxetaCD}
\end{minipage}
\hspace{\fill}
\begin{minipage}[t]{75mm}
\hspace*{3mm}
\includegraphics*[scale=0.35]{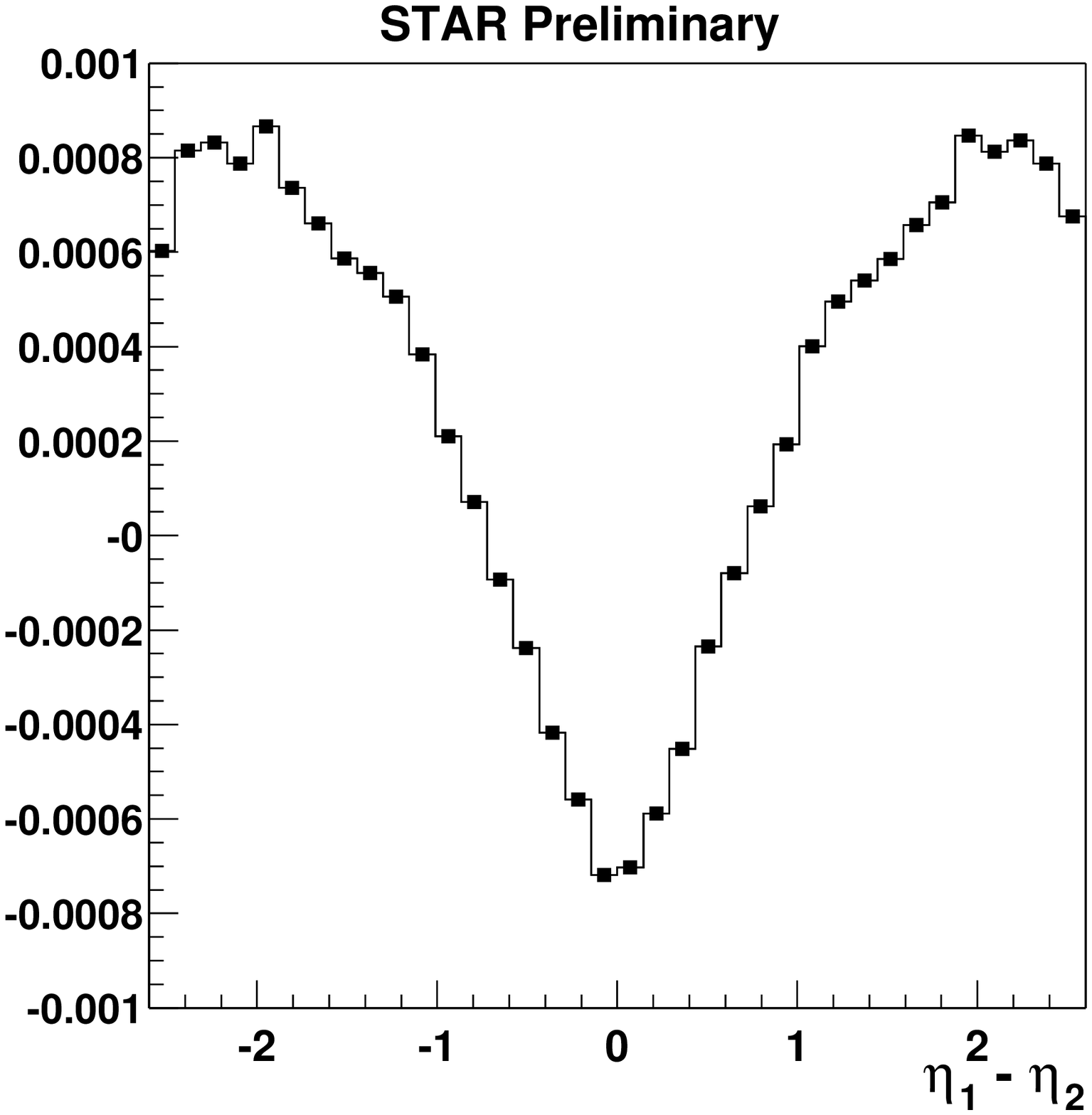}
\vspace{-10mm}
\caption{CD autocorrelation projection onto the ($\eta_1 - \eta_2$) axis for
$\sqrt{s_{NN}} = 130$~GeV Au+Au collisions from Fig.~\ref{Fig:etaxetaCD}.}
\label{Fig:etadiffauto}
\end{minipage}
\end{figure}

Correlations in pseudorapidity ($\eta$) space are affected by string
fragmentation, source expansion, medium effects, etc.
Fig.~\ref{Fig:etaxetaCD} displays, for example, the two-particle
CD pseudorapidity correlation $\eta_1 \otimes \eta_2$,
for 130~GeV Au+Au
collision data from STAR for central events (multiplicities $> 0.81$ of
maximum) using primary
particles with $0.15 \leq p_T \leq 2$~GeV/c, $|\eta| < 1.3$, and full azimuthal
acceptance. 
The charge-pair specific correlations were obtained by dividing the
number of particle pairs of each charge combination
from the same event (sibling pairs)
in a given bin (0.1 in width), summed over all events in the sample, by the 
number of corresponding charged particle mixed-pairs in the bin
from different events.
The latter events were required to have
primary vertex locations within 5~cm in the TPC along the beam axis. 
The correlation function was normalized
based on the total number of pairs used in the numerator and denominator.
Large-scale structure covering the full STAR
pseudorapidity acceptance is evident in Fig.~\ref{Fig:etaxetaCD}
where the dark band along the
$\eta_1 = \eta_2$ diagonal is negative
and the off-diagonal corners are positive.
The dominant structure is with respect to the difference variable
$\eta_1 - \eta_2$.
Projection onto the $\eta_1 - \eta_2$ axis,
shown in Fig.~\ref{Fig:etadiffauto}, results
in the CD {\it autocorrelation}.
This is one example of long-range axial (longitudinal or azimuthal) 
correlations which potentially carry new information about the dynamics
and evolution of the collision.

\begin{figure}[htb]
\begin{minipage}[t]{80mm}
\includegraphics*[scale=0.35]{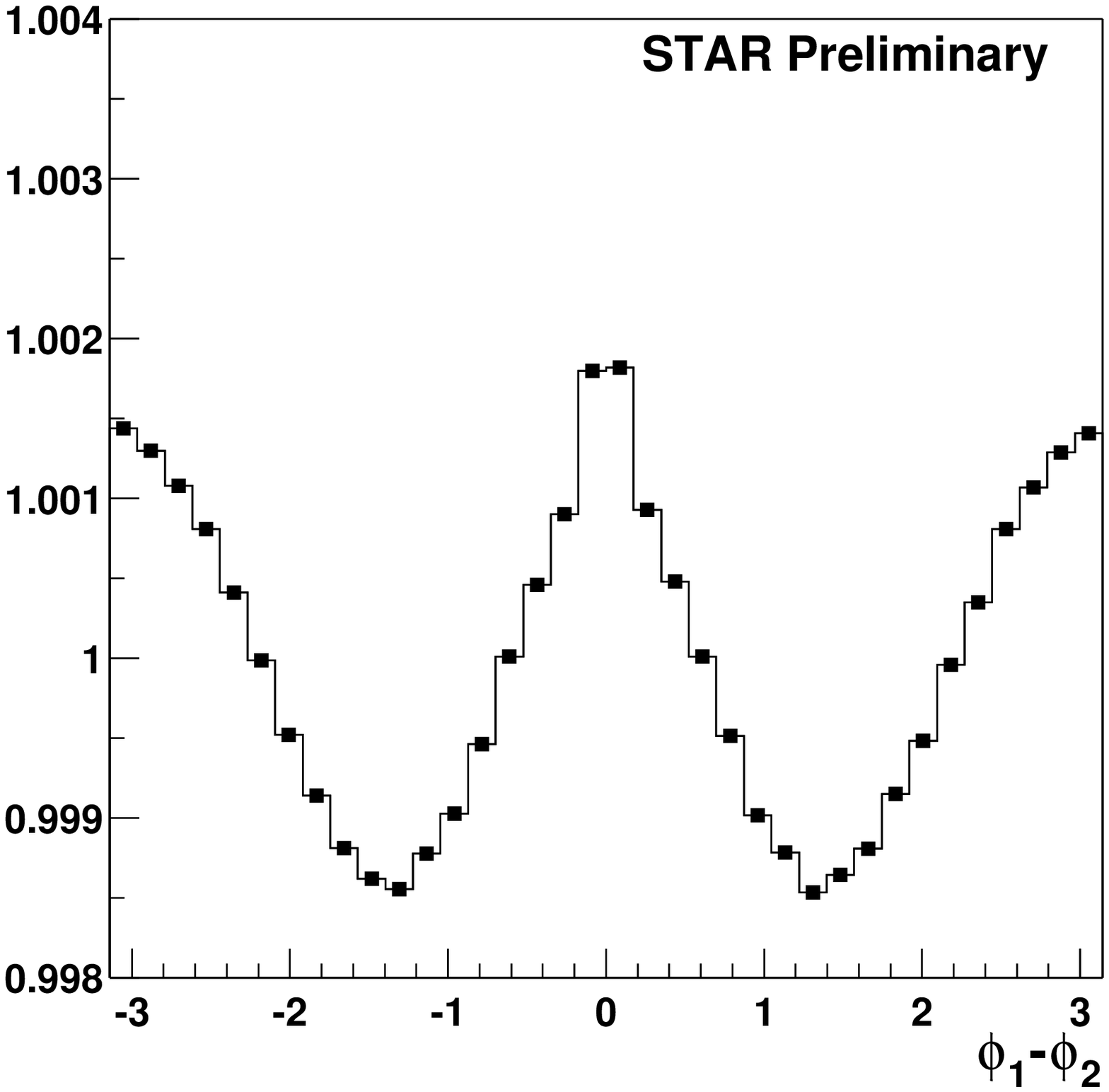}
\vspace{-10mm}
\caption{Autocorrelation projection onto the
$(\phi_1 - \phi_2)$ axis in radians for like-sign charged particle pairs for
$\sqrt{s_{NN}} = 130$~GeV Au+Au collisions.}
\label{Fig:philikediffauto}
\end{minipage}
\hspace{\fill}
\begin{minipage}[t]{75mm}
\hspace*{5mm}
\includegraphics*[scale=0.35]{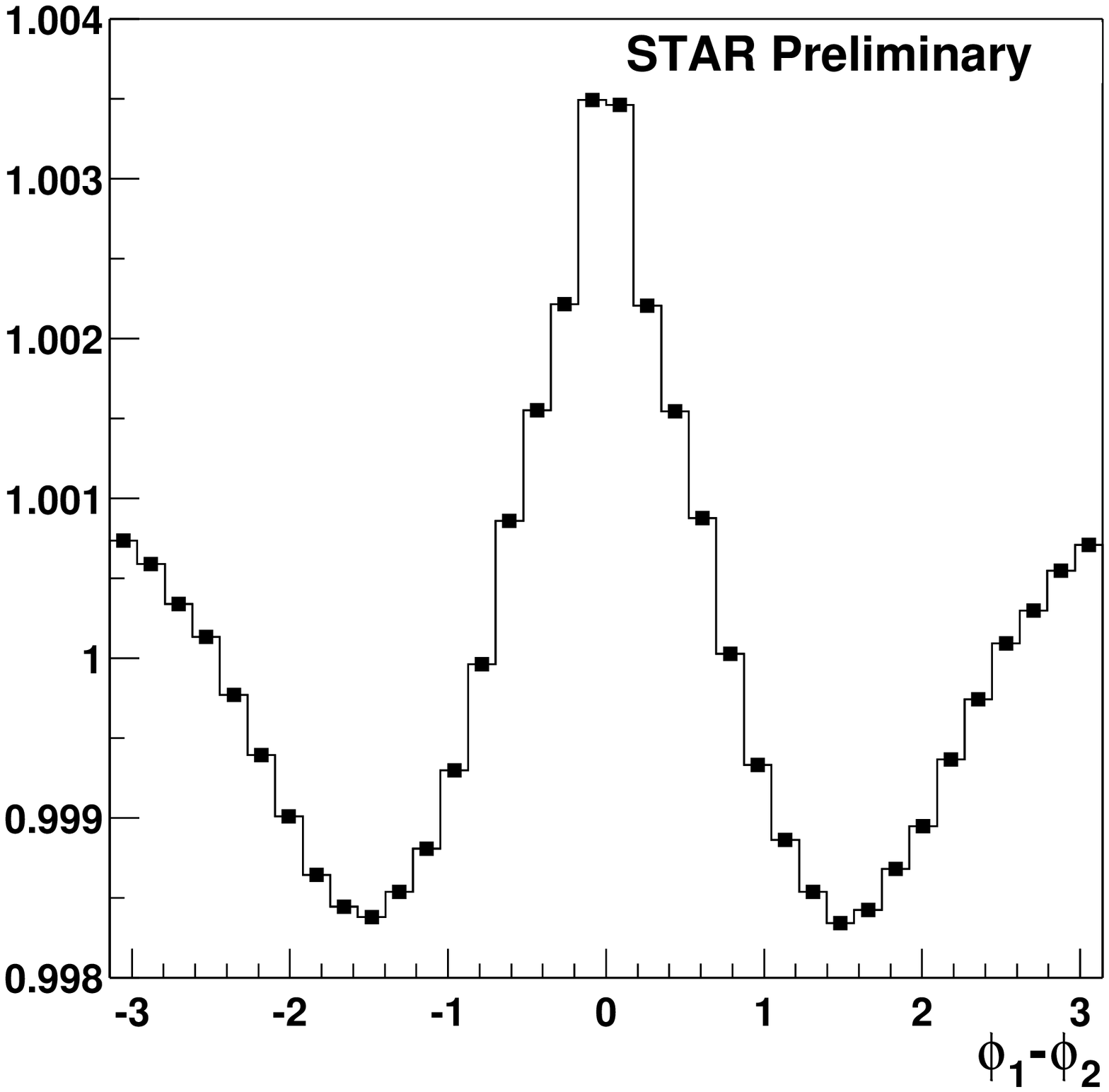}
\vspace{-15mm}
\caption{Autocorrelation projection onto the
$(\phi_1 - \phi_2)$ axis in radians for unlike-sign charged particle pairs for 
$\sqrt{s_{NN}} = 130$~GeV Au+Au collisions.}
\label{Fig:phiunlikediffauto}
\end{minipage}
\end{figure}

\section{Long-Range Azimuthal Angular Correlations}

Two-dimensional azimuthal angle correlations, $\phi_1 \otimes \phi_2$,
are dominated by elliptic flow contributions which produce a
$1+2v_2^2 \cos [2(\phi_1 - \phi_2)]$ dependence.  Autocorrelation projections
onto the $\phi_1 - \phi_2$ axis are shown in Figs.~\ref{Fig:philikediffauto}
and \ref{Fig:phiunlikediffauto} for like and
unlike charge-sign pairs for $\sqrt{s_{NN}} = 130$~GeV Au+Au collision data
from STAR for combined mid-central and central events
(multiplicities greater than 0.47 of maximum)
assuming the same primary track acceptances described in the preceding section.
In addition to elliptic flow, this kind of analysis
reveals long-range non-flow components~\cite{starflowprc}
({\it e.g.} non-collective, few-body processes such as resonance
decays, mini-jets, etc.) which appear as the $non-\cos [2(\phi_1 - \phi_2)]$
dependences in Figs.~\ref{Fig:philikediffauto} and \ref{Fig:phiunlikediffauto}.

A possible method to assess the importance of non-flow components
has been discussed by Borghini {\it et al.}~\cite{borghini}. It is based on
the technique of forming four-particle cumulants
for studying azimuthal
an\-i\-so\-tro\-py in particle production from heavy ion collisions.
In the cumulant expansion, dependences on lower-order
correlation contributions are eliminated term-by-term
leaving only those contributions
arising from true, four-particle correlations.  In this way, the
deduced values of $v_2$ obtained with this procedure
are expected to more accurately measure the
collective, elliptic flow produced in heavy ion collisions.

\begin{figure}[htb]
\begin{minipage}[t]{80mm}
\includegraphics*[scale=0.4]{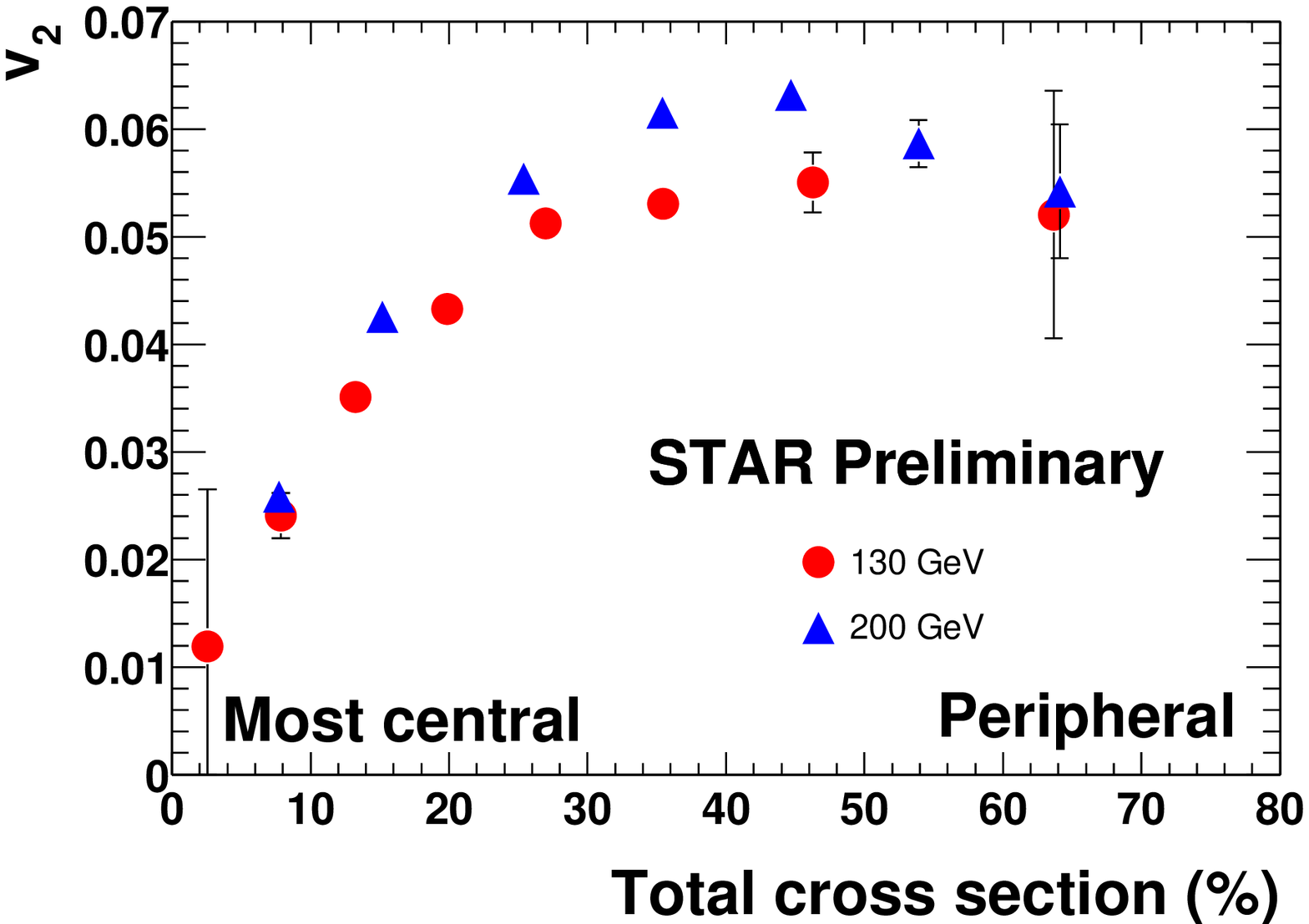}
\vspace{-10mm}
\caption{Preliminary azimuthal anisotropy measurements of
$v_2$ versus centrality for nonidentified charged particles
using four-particle cumulants
for $\sqrt{s_{NN}} = 130$ and 200~GeV Au+Au collisions.}
\label{Fig:v2centrality}
\end{minipage}
\hspace{\fill}
\begin{minipage}[t]{75mm}
\includegraphics*[scale=0.4]{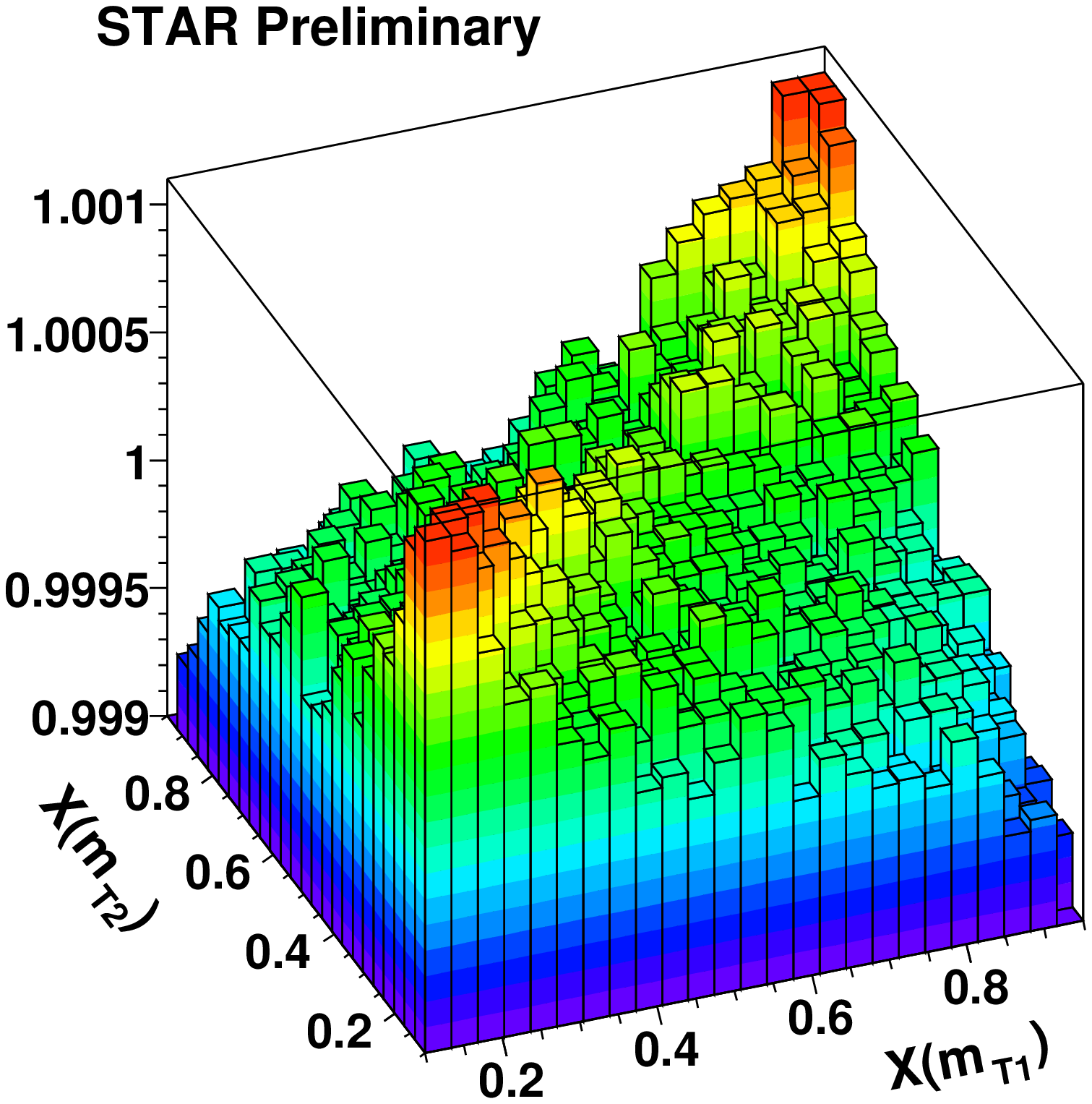}
\vspace{-10mm}
\caption{Perspective view of the two-dimensional $X(m_{T1}) \otimes X(m_{T2})$
charge independent correlations for $\sqrt{s_{NN}} = 130$~GeV Au+Au
collision data.} 
\label{Fig:mt1xmt2}
\end{minipage}
\end{figure}

This new method was implemented in
STAR~\cite{starflowprc} and used to analyze the 
$\sqrt{s_{NN}} = 130$ and 200~GeV Au+Au data. 
Values of $v_2$ obtained in this way
are reduced about 15\% relative to that obtained with either the standard
reaction plane method~\cite{starflowprl}
or with two-particle angular correlations~\cite{starflowprc}.
A comparison of $v_2$ versus centrality using 
four-particle cumulants for the $\sqrt{s_{NN}} = 130$
and 200~GeV Au+Au data is shown in
Fig.~\ref{Fig:v2centrality} where it is noted that
$v_2$ on average increases about 7-10\%
at the higher energy.

From the study of elliptic flow in STAR it has been found that the event
plane resolution is sufficient to conduct
three-dimensional Pratt-Bertsch HBT analysis with respect to event plane.
This type of analysis of charged pion data for $\sqrt{s_{NN}} = 130$ and 200~GeV
Au+Au collisions was reported in
these proceedings
by M. Lopez-Noriega.  This analysis indicates an out-of-event-plane extended
source with $\sim 5$\% azimuthal anisotropy
for the geometrical source shape.  The
azimuthal dependence of the HBT radii {\it cannot}
be explained
assuming {\it only} azimuthal momentum anisotropy or a source with
{\it only} azimuthal anisotropy in the density.  In particular, the blast-wave
model provides consistent descriptions of both
$v_2(m,p_T)$ and the azimuthally dependent HBT radii.  Interestingly,
the azimuthal source shape anisotropies at
RHIC and the AGS~\cite{lisae895} are both
out-of-plane extended but the magnitude at RHIC is about one-tenth
as large.

\section{Long-Range Transverse Momentum Correlations}

The shape of the transverse momentum spectrum of charged hadrons near
mid-rapidity produced in relativistic heavy ion collisions is affected by
quantum statistics, thermalization, transverse expansion, hard QCD processes,
mini-jets, etc.  In addition, non-statistical or dynamical fluctuations
within each event or from
event-to-event modify the shape of the inclusive $p_T$ distribution,
increasing the yield at higher $p_T$. 
For example, an intrinsic Boltzmann distribution
with gaussian-like event-by-event fluctuations in the inverse
temperature parameter becomes a L\'{e}vy
distribution in the inclusive $p_T$ spectrum~\cite{clt}.
Similarly, fluctuations in inverse
temperature produce a long-range
``saddle shape'' in the two-dimensional $m_{T1} \otimes m_{T2}$ correlation
function~\cite{clt}, which is the
transverse momentum space analog of the two-dimensional
correlations shown in the preceding sections. 
This is shown in Fig.~\ref{Fig:mt1xmt2} for the
5\% most central Au+Au events at $\sqrt{s_{NN}} =130$~GeV
using all charged particles
with $0.15 \leq p_T \leq 2.0$~GeV/$c$, $|\eta| < 1.3$, and full $2\pi$
acceptance.\footnote{Approximately constant statistics in each bin were
achieved by mapping $p_T$ into
$X(m_T)$
where $X(m_T) = 1 - [(1 + m_T/T) \exp (-m_T/T)]/[(1 + m_0/T) \exp (-m_0/T)]$;
$m_0$ = pion mass was assumed.}
The dominant features in these data
include HBT and Coulomb
interaction correlations which produce the
ridge at low $m_T$, the peak at large momentum,
plus the long-range saddle shape.
All features have similar amplitudes and are the result of non-statistical
processes. 

Further evidence that non-statistical
event-by-event fluctuations occur in RHIC collisions is
observed in the distribution of event-wise average $p_T$
($\langle p_T \rangle$).  
From application of the central limit theorem (CLT)~\cite{clt,tannenbaum}
it is known that in the
absence of non-statistical fluctuations, the mean and $rms$ width of the
event-wise $\langle p_T \rangle$ distribution are equal to 
$\overline{p_T}$ and $\sigma_{\hat{p}_T}/\sqrt{N}$, respectively, where
$\overline{p_T}$ and $\sigma_{\hat{p}_T}^2$ are the mean and variance of the
inclusive $p_T$ distribution and $N$ is the event multiplicity. 
Increased width in the data beyond this amount provides further evidence for 
non-statistical fluctuations. 

The inclusive $p_T$
spectrum of charged particles near mid-rapidity over the $p_T$ range of
interest here is described very well
by a Gamma distribution~\cite{tannenbaum}. 
The $\langle p_T \rangle$ distribution
for RHIC events which would be expected
in the absence of non-statistical fluctuations is accurately described
by the $N$-fold convolution of the inclusive $p_T$ Gamma distribution.
This is also 
a Gamma distribution~\cite{tannenbaum} whose variance is in agreement with 
the CLT~\cite{clt,tannenbaum}.  The
accuracy of the $N$-fold Gamma distribution
for representing the statistical-fluctuations-only
$\langle p_T \rangle$ reference distribution
is assured, for moderate to large values of $N$,
by the strong suppression
of higher-order moments in this distribution.  This suppression
is driven by powers of $N$ which increase with moment number.
Traditional mixed event and Monte Carlo sampled references were also
obtained and shown to be
consistent with the more accurate Gamma distribution reference.

\begin{figure}[htb]
\begin{minipage}[t]{70mm}
\hspace*{3mm}
\includegraphics*[scale=0.6]{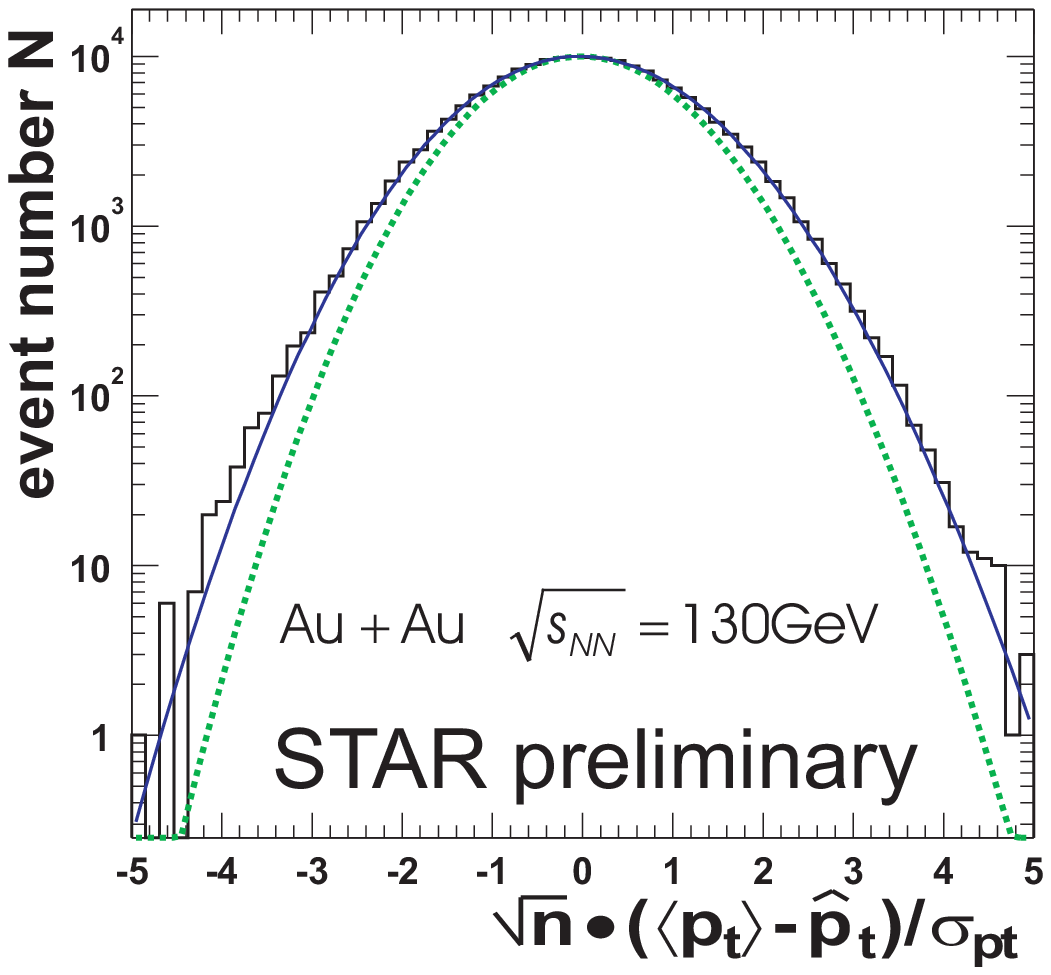}
\vspace{-10mm}
\caption{Mean $p_T$ distribution for $\sqrt{s_{NN}} = 130$~GeV
Au+Au central collisions
with respect to $\overline{p_T}$ in units of
$\sigma_{\hat{p}_T}/\sqrt{N}$ compared to the Gamma distribution reference
expected in the absence of non-statistical fluctuations
(dotted curve) and a Gamma distribution calculated with
$rms$ width increased by 14\%
(solid curve).}
\label{Fig:meanptdist}
\end{minipage}
\hspace{\fill}
\begin{minipage}[t]{85mm}
\includegraphics*[scale=0.5]{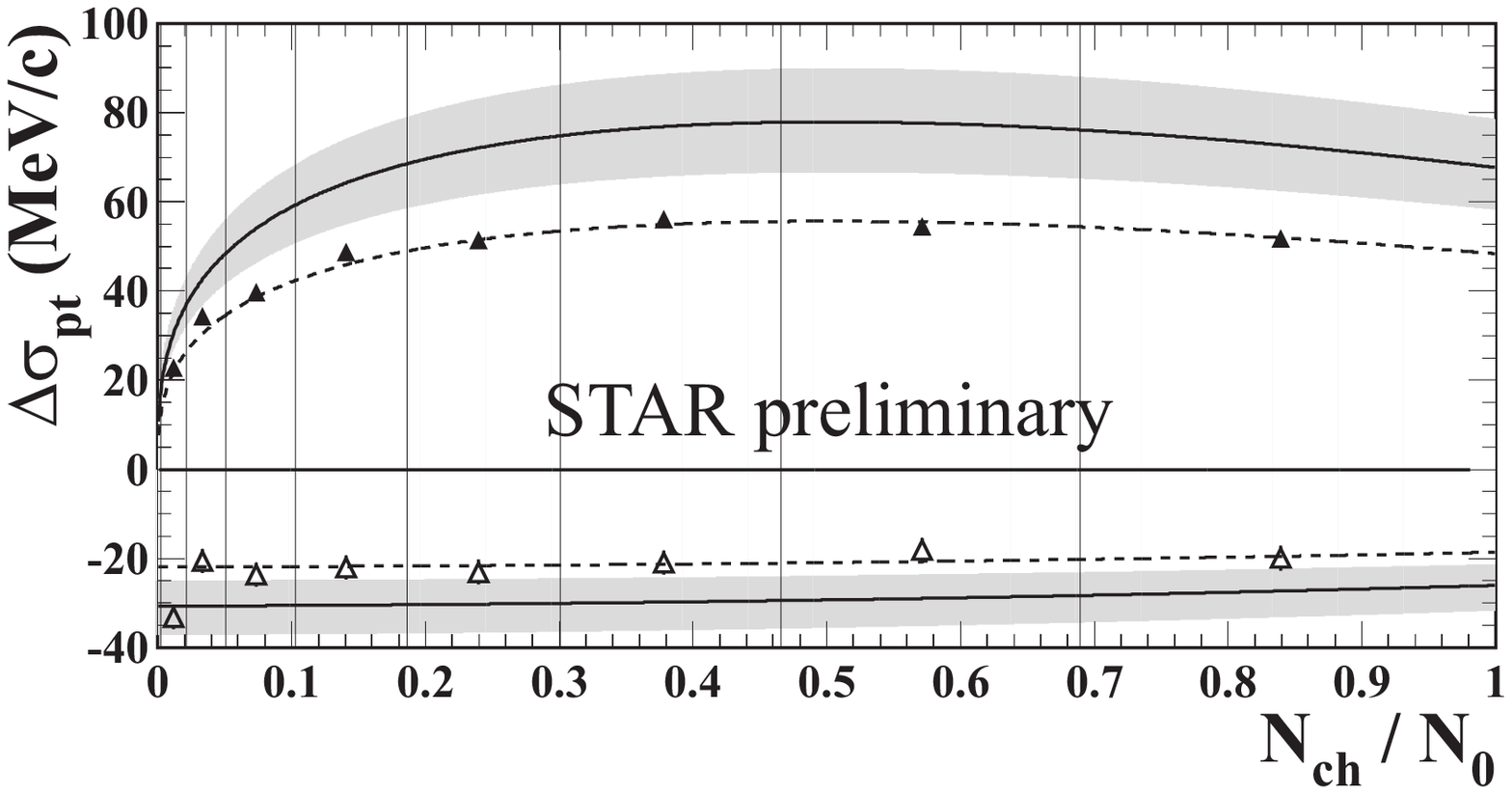}
\vspace{-10mm}
\caption{Mean $p_T$ CI (solid triangles) and CD (open triangles)
fluctuation measures
for 130~GeV minimum bias Au+Au events versus centrality as explained in the
text.  The CD quantities have been multiplied by 3 for clarity.}
\label{Fig:delsigpt}
\end{minipage}
\end{figure}

The frequency distribution\footnote{The distribution shown is for the
quantity $[\langle p_T \rangle - \overline{p_T}]/[\sigma_{\hat{p}_T}
/\sqrt{N}]$ which is designed to suppress the slight dependence of
$\overline{p_T}$ on $N$, remove the $1/\sqrt{N}$ statistical broadening, and
eliminate dependence on the variance of the inclusive distribution.}
of event-wise $\langle p_T \rangle$ for the upper
15\% central Au+Au collisions at $\sqrt{s_{NN}} =130$~GeV from STAR is shown
in Fig.~\ref{Fig:meanptdist} in comparison with
the Gamma distribution reference.  These data
are from 183K events and use 70\% of all primary charged particles 
(less than 100\% due to tracking inefficiency
and track quality requirements) with
$0.1 \leq p_T \leq 2.0$~GeV/$c$, $|\eta| < 1$, and full $2\pi$
acceptance.  Striking evidence is observed for large, non-statistical
fluctuations which increase the width of the measured distribution by
14$\pm$0.5(stat)\% relative to the Gamma distribution reference.

Quantitative measures of $\langle p_T \rangle$ non-statistical fluctuations
are provided by a number of quantities~\cite{reid,phipt,s2d}, for example:
\begin{eqnarray}
\Delta\sigma^2_{p_T} & \equiv & \frac{1}{\varepsilon} \sum_{j=1}^{\varepsilon}
N_j \left( \langle p_T \rangle_j - \overline{p_T} \right)^2
- \sigma_{\hat{p}_T}^2
\equiv 2 \sigma_{\hat{p}_T} \Delta\sigma_{p_T}, \\
\Phi_{p_T} & \equiv & \left[ \frac{1}{\varepsilon} \sum_{j=1}^{\varepsilon}
\frac{N^2_j}{\langle N \rangle} ( \langle p_T \rangle_j - \overline{p_T} )^2 
 \right]^{1/2} - \sigma_{\hat{p}_T}, \\
\sigma^2_{\langle p_T \rangle,{\rm dynamical}} & \equiv &
\frac{1}{\varepsilon} \sum_{j=1}^{\varepsilon}
\frac{1}{N_j (N_j - 1)}
\sum_{i \neq i^{\prime} = 1}^{N_j} \delta p_{T_{ji}} \delta p_{T_{ji^{\prime}}},
\end{eqnarray}
where $\varepsilon$ is the number of events, $j$ is the event index,
$N_j$ is the event multiplicity, $\langle N \rangle$ is the mean multiplicity,
$i$ is a particle index, and $\delta p_{T_{ji}} = p_{T_{ji}} - \overline{p_T}$.
For minimal variations of $N_j$ within the event ensemble
\begin{eqnarray}
\Delta\sigma_{p_T} & \cong & \Phi_{p_T} \cong \frac{\langle N \rangle - 1}
{2 \sigma_{\hat{p}_T}} \sigma^2_{\langle p_T \rangle,{\rm dynamical}}.
\end{eqnarray}
Charge sum (CI) and difference (CD) measures corresponding to
$\Delta\sigma_{p_T}$ are given by
\begin{eqnarray}
\langle N \rangle \Delta\sigma^2_{p_T,CI/CD} & \equiv &
\langle N_+ \rangle \Delta\sigma^2_{p_T++} +
\langle N_- \rangle \Delta\sigma^2_{p_T--} \pm
2 \sqrt{\langle N_+ \rangle \langle N_- \rangle}
\Delta\sigma^2_{p_T+-},
\end{eqnarray}
where $\Delta\sigma^2_{p_T \pm \pm} = 2 \sigma_{\hat{p}_T\pm}
\Delta\sigma_{p_T \pm \pm}$, $\Delta\sigma^2_{p_T+-} = 
2 \sqrt{ \sigma_{\hat{p}_T+} \sigma_{\hat{p}_T-} }
\Delta\sigma_{p_T+-}$, and 
$\Delta\sigma^2_{p_T+-}$ is evaluated noting that the covariance,
$\sigma^2_{\hat{p}_T+,\hat{p}_T-}$, vanishes via the CLT.

The centrality dependence of $\Delta\sigma_{p_T}$
is shown in Fig.~\ref{Fig:delsigpt}
for 205K $\sqrt{s_{NN}} =130$~GeV Au+Au minimum bias
events from STAR using 70\% of all charged primary particles for charge
independent and charge dependent measures.
Statistical errors are $\pm 0.5$~MeV/$c$.  The dashed lines are polynomial fits
to the data, the solid lines are estimated extrapolations to 100\% of the
primary particles in the acceptance, and the
shaded bands indicate $\pm 15$\% systematic errors.

These data provide a quantitative measure of the non-statistical
fluctuations visually demonstrated in Fig.~\ref{Fig:meanptdist}
and reveal an
intriguing dependence on centrality. They also show that the CI (CD) 
fluctuations at RHIC are larger (smaller) than at the SPS~\cite{spsmeanpt}. 
The PHENIX
experiment's null result~\cite{phenixmeanpt},
which was carried out with much smaller
$\eta,\phi$ acceptance than in STAR, does not contradict STAR's observation
of strong, non-statistical fluctuations.

%

\section{Summary and Conclusions}

During the first two years of RHIC operation the STAR experiment has
produced a wealth of new correlation, flow, and event-by-event fluctuation
measurements.\footnote{Event-by-event fluctuations in net charge were also
measured for the 130 and 200~GeV Au+Au and p+p collision data; see
the contribution to these proceedings by C. Pruneau.} 
%
The results for the $\sqrt{s_{NN}} = 200$~GeV Au+Au collisions are similar to
those for the 130~GeV data.  Compared to
heavy ion results at lower energies from the AGS and SPS, the RHIC events
display (1) similar HBT correlation lengths, (2) $R_o/R_s \approx 1$,
(3) larger pion phase space densities, (4) stronger elliptic flow $v_2$,
(5) larger (smaller) charge independent (dependent) mean $p_T$ fluctuations,
(6) strong transverse expansion with (apparent)
brief emission duration, and (7) similar
out-of-event-plane extended sources but with much smaller azimuthal source
shape anisotropies.  The STAR measurements, the successful
description of the data in terms of the blast-wave model, and the
general success of the hydrodynamical approaches~\cite{hydro}
at RHIC energies, taken together, suggest
that the Au+Au collisions at RHIC are governed by rapid thermal equilibration,
very high initial 
pressures, followed by hydrodynamical expansion.
Analyses of the emerging fluctuation measurements from STAR will also
help determine the source structure and dynamics
and allow further differentiation among
QGP and purely hadronic scenarios.

%

\end{document}